\documentclass[%
 preprint,
 amsmath,amssymb,
 aps,
]{revtex4-1}

\usepackage{graphicx}
\usepackage{dcolumn}
\usepackage{bm}

\begin{document}

\title{Modeling a Double-Spending Detection System for the Bitcoin Network}

\author{Marco Alberto Javarone}
\email{marcojavarone@gmail.com}
 \affiliation{nChain Ltd, London, UK}
 \affiliation{School of Computing, Electronics and Mathematics, Coventry University, Coventry, UK}
 
\author{Craig Steven Wright}%
 \email{c.wright@nchain.com}
\affiliation{nChain Ltd, London, UK}

\date{\today}

\begin{abstract}
The Bitcoin protocol prevents the occurrence of double-spending (DS), i.e. the utilization of the same currency unit more than once.
At the same time a DS attack, where more conflicting transactions are generated, might be performed to defraud a user, e.g. a merchant.
Therefore, in this work, we propose a model for detecting the presence of conflicting transactions by means of an 'oracle' that polls a subset of nodes of the Bitcoin network. We assume that the latter has a complex structure. So, we investigate the relation between the topology of several complex networks and the optimal amount, and distribution, of a subset of nodes chosen by the oracle for polling.
Results show that small-world networks require to poll a smaller amount of nodes than regular networks. In addition, in random topologies, a small number of polled nodes can make a detection system fast and reliable even if the underlying network grows.
\end{abstract}

\maketitle

\section{Introduction}
Nowadays, the Blockchain~\cite{satoshi01} and the world of cryptocurrencies represent a vibrant area in industry and academia. Notably, new studies (e.g.~\cite{barronchelli01,barronchelli02,suweil01,javarone02,vattar01}) are enriching the scientific literature and new services, based on this technology, are pervading different sectors~\cite{block01}.
In few words, the Blockchain can be described as a distributed ledger composed of 'blocks' containing a list of transactions. Its structure is simple, i.e. its blocks connect generating a growing chain. Notably, each new block connects to the head of the chain.
Users can easily generate a new block by collecting a set of transactions (not yet recorded in the previous blocks). However, adding a new block to the Blockchain is all but easy, and many users (called 'miners') take part in this process, in a challenging competition. 
Notably, the first miner that completes a process of 'Proof-Of-Work' (POW)~\cite{antinopolis01} succeeds and can add her/his block to the Blockchain. Then, miners move to the formation of a new block, that will be connected to the latest 'discovered' block (i.e. the one on the head of the chain). The POW requires the availability of computational power, only the successful miner (i.e. the first one that completes the POW) receives a reward in Bitcoin~\cite{satoshi01}.
It is worth to highlight that more than one block can be 'discovered' (by performing the POW) at the same time. So, occasionally a 'fork' can emerge in the Blockchain. However, forks are easily solved by the system since, according to the spreading dynamics in the Bitcoin network and to the possibility to connect a new block to only one head of the fork, one branch will soon become longer than the others. Thus, according to the protocol, only the longest branch will be considered by miners, while the others will be cut.
Looking at the recent dynamics of the cryptocurrency market (see for instance~\cite{barronchelli01}), we can appreciate the high impact that Bitcoin is experiencing also at societal level. However, like any other digitalized system, it can be subjected to various attacks. 
It is important to highlight that double-spending (DS), i.e. the fraudulent attempt to use the same Bitcoin more than once, cannot succeed in the Bitcoin network~\cite{satoshi01,antinopolis01}. However, the generation of conflicting transactions, following the basic mechanism of a DS attack, can be adopted to defraud users. 
Therefore, the detection of DS attempts is of paramount relevance for increasing the safety of users, like merchants, that are willing to accept payment in Bitcoin. 
A complete knowledge of the structure of the Bitcoin network would be extremely useful for the design of a system that identifies a promptly reports DS attacks. In addition, even by collecting data related to a fraction of the network to infer its topological structure~\cite{javarone02}, we have to consider that the latter can evolve over time, e.g. new users can join the network and new connections can be generated. As result, only a limited information can be actually available. Furthermore, the early phases of a DS attack can involve each time different nodes of the network.
Beyond that, it is worth to remind that a central core of nodes is highly connected, i.e. the set of miners.
For these reasons, we suggest that a reliable DS detection system can be modeled by using the framework of the modern theory of networks~\cite{estrada01}, exploiting in particular the core composed of miners. Thus, assuming that the Bitcoin network has a non-trivial structure, we aim to investigate the performance of an oracle that polls a subset of nodes in order to verify the presence of conflicting transactions. With this goal in mind, we focus on relation between the topology of different complex networks and the definition of a random subset of nodes to be polled for the detection task.
Since the analysis considers global properties of a network, we deem that the related results can be useful for implementing a DS detection system in the Bitcoin network. 
Accordingly, the target of our investigation is estimating the optimal amount of polled nodes for a rapid detection of DS attacks. Notably, we analyze different complex topologies, and a random distribution of polled nodes over each network.
The reminder of the paper is organized as follows: Section~\ref{sec:double-spending} provides a more detailed description of Bitcoin transactions and the mechanism of a DS. Section~\ref{sec:model} introduces the proposed model. Section~\ref{sec:results} shows results of numerical simulations. Eventually, Section~\ref{sec:conclusion} ends the paper.
\section{Bitcoin transactions}\label{sec:double-spending}
In this section, we provide a more detailed description of Bitcoin transactions~\cite{antinopolis01}, then we focus on the basic mechanism of a DS attack, performed by the generation of conflicting transactions.
Transactions are described by different parameters, e.g. keys, signatures, and so on, and contain one or more inputs and, in general, one or two outputs.
A set of inputs is used to define the amount of Bitcoin of a transaction, which has the related outputs, indicating where these Bitcoin are sent.
Accordingly, users can spend Bitcoin accumulated during previous transactions. For instance, a user A has a total of $6$ Bitcoin collected during two transactions, i.e. one whose value was $2$ Bitcoin and another one of $4$ Bitcoin. 
Now, if the user A has to send $5$ Bitcoin to the user B, can define a transaction whose input corresponds to the set of transactions previously received. Then, in this case, the new transaction has two outputs: one whose value is equal to $5$, sent to B, and another one directed toward herself/himself (i.e. the A's wallet) equal to (or smaller than) $1$. If the value of the second output is smaller than $1$, the difference constitutes the 'fee' claimed by the winning miner (fee $=$ input $-$ output).
Thus, after the described transaction, A has only (or less than) $1$ Bitcoin, while B can now spend the received $5$ Bitcoin. 
Transactions are broadcast through the Bitcoin network, to be included as soon as possible into the Blockchain. At the beginning of this process, they are unverified, i.e. they are defined as $0$-confirmation transactions.
Here, a fraudulent user could try to perform a DS. As above reported, it is not possible to spend twice the same Bitcoin. So, a DS attack to the network fails.
However, here we focus on the utilization of a DS attack to defraud another user. Notably, let us now consider the presence of a third user, say C, that cooperates with A. As before, the latter performs a first transaction sending Bitcoin to her/his accomplice (i.e. user C). Then, few instant later, A performs a second transaction addressed to B. If the inputs of the two transactions correspond, completely or only in part (i.e. a subset of input), A is attempting a fraud. 
The same, in principle, applies if the first transaction is addressed to user B, and few instant later a second one (always using same input) is addressed to user C.
During a DS attack, two (or more) conflicting transactions (i.e. that share a subset of the input) spread over the network, and only one will be added into the Blockchain.
Considering the previous example, if the recorded transaction is the honest one (i.e. that sent to B) the fraud fails whereas, in the opposite case, A and C accomplish their goal, and B does not receive her/his $5$ Bitcoin.
A pictorial representation of this scenario is provided on the top of Figure~\ref{fig:figure_1}.
\section{Model}\label{sec:model}
\begin{figure}[h!]
\centering
\includegraphics[width=0.8\textwidth]{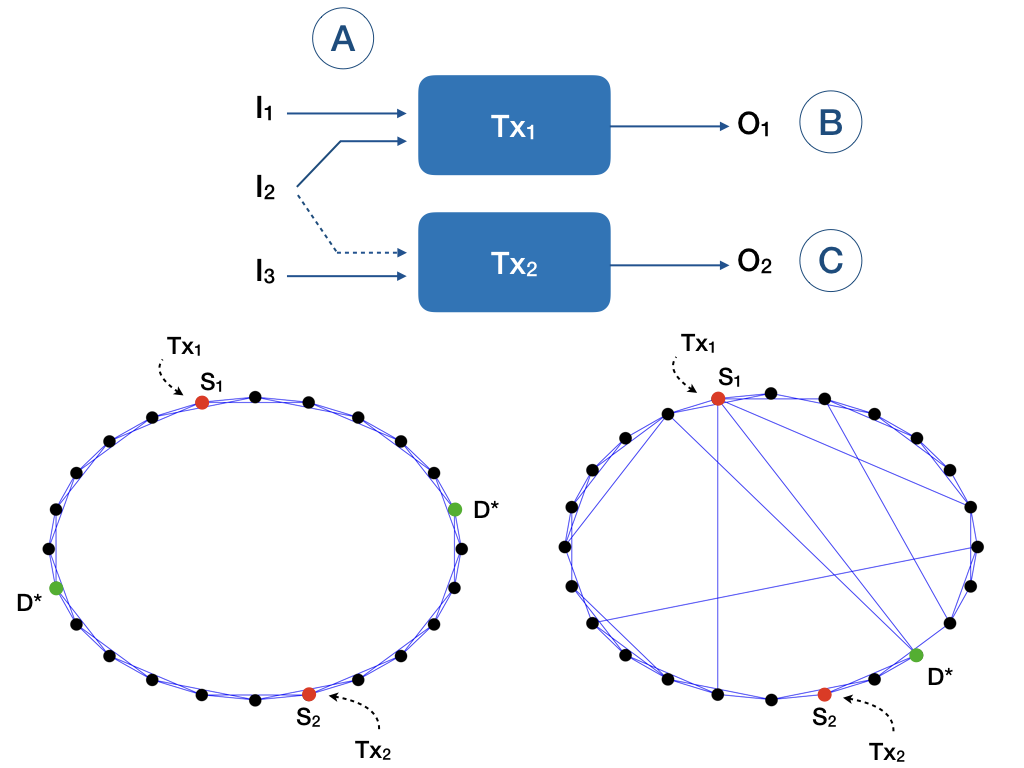}
\caption{Pictorial representation of a DS attack and its broadcasting through two different networks, i.e. a regular ring (on the left) and a small-world network (on the right). On the top, the schematic illustration of two transactions $Tx_{1}$ and $Tx_{2}$, generated by the user A, and sent to user B and user C, respectively. The inputs of the two transactions partially coincide, since the input $I_2$ is used twice, i.e. user A is attempting a DS attack.\label{fig:figure_1}}
\end{figure}

As above described, transactions are broadcast into the Bitcoin network, then (if valid) are recorded into the Blockchain. For the sake of simplicity, broadcasting a transaction can be modeled as a spreading process. A wide number of investigations analyzed these dynamics in complex networks, spanning from social dynamics~\cite{loreto01,liu01,javarone01} to epidemic processes~\cite{vespignani01,boccaletti01}. Therefore, a similar approach can be adopted also for studying our problem.
At the same time, it is worth to remind that the detection of a DS attack is a bit different from a classical spreading process, e.g. epidemic dynamics. 
For instance, in viral spreading we can measure the amount of time steps a virus takes for infecting all nodes of a network, on varying different parameters. In a similar way, in social dynamics we can study the amount of time steps required for sharing information through a social media, and so on and so forth. Instead, in our case, we need to consider the dynamics of two (or even more) spreading processes to evaluate the time required for their detection by means of a polling. It is important to mention that miners cannot be aware whether a transaction results from a fraudulent attack and, in addition, they are 'rational'~\cite{javarone03}, i.e. they only aim to generate blocks collecting transactions that release a high fee.
We remind that a DS attack entails that two, or more, conflicting transactions are broadcast.
However, the higher the amount of these conflicting transactions, the shorter the time required to detect a fraudulent attempt (as only two of them are detected, the fraud is discovered). So, the worse case is based on the presence of only two conflicting transactions. For that reason, the proposed model considers only that case.
Furthermore, for the sake of simplicity, our fraudulent user broadcasts the two transactions only to two different nodes in the Bitcoin network (although, in the real system, nodes have usually more than one neighbor). Then, the two selected nodes play the role of source or, using the epidemilogist jargon, they are a sort of 'patient-$0$'.
Since we aim to study the relation between the topology of a network and the amount of polled nodes, for an early detection, our investigation focuses on the time our oracle should wait before performing the poll on a randomly defined set of nodes. Here, the time is computed in terms of time steps (the actual equivalence between 'time' and 'time steps' requires further details not provided in this work).
The waiting time considered by our model allows to satisfy both scenarios related to the fraudulent case above described, i.e. when the malicious transaction is performed before than the honest one, and when it is performed a bit later. In both cases, we assume the the temporal lag between the two transaction is extremely small.
In addition, the concept of time step and that of distance are strongly connected. Notably, given two nodes, say $x$ and $y$, the amount of time steps required for sending a transaction from $x$ to $y$ corresponds to the amount of links a transaction has to go through for moving from node $x$ to node $y$.
Thus, since the oracle needs both transactions, the amount of time steps required is equal to the distance between a polled node and the farer source (i.e. the farer patient-$0$).
Summarizing, the proposed model is then defined as follows: we generate a network with $N$ nodes; two of them are randomly selected as sources (i.e. those receiving the two transactions at $t=0$), and $Z$ nodes are randomly chosen to perform a poll. 
In doing so, with a very low probability, a node can be both a source and a polled node at the same time.
Then, we compute the amount of time steps required by the set of the $Z$ polled nodes for receiving both transactions (i.e. the system needs that only one of these nodes receives both transactions).
Figure~\ref{fig:figure_1} shows the basic configuration of the model in two different networks (i.e. a regular ring on the left and a small-world network on the right).
As we know from previous investigations (e.g.~\cite{vespignani02,lu01}), small-world networks and scale-free networks support spreading processes. In particular, small-world networks have an average shortest-path length that scales with the logarithm of the size of a network (i.e. the shortest distance between two randomly chosen nodes $x$ and $y$ is, on average, $d_{x,y} \sim \ln(N)$, with $N$ number of nodes. Instead, scale-free networks can be even faster than small-world networks, in spreading processes, given the presence of highly connected hubs (i.e. nodes with a huge amount of connections).
For that reason, both small-world and scale-free networks are expected to support a quick detection of conflicting transactions.
The pictorial representation in Figure~\ref{fig:figure_1} shows two networks with red nodes, i.e. sources $s_1$ and $s_2$, and green nodes, labeled by $D^*$, to indicate polled nodes. In both networks, polled nodes occupy an optimal position (i.e. as a transaction spreads from $s_1$ and one from $s_2$, polled nodes are located in positions that ensure the fastest detection).
It is interesting to note that, according to the topology of a network and to the amount its nodes, given the position of sources (i.e. $s_1$ and $s_2$), the optimal locations for a quick detection might vary case by case. For instance, the regular ring on the left-hand side of Figure~\ref{fig:figure_1} indicates the presence of two optimal nodes, while the small-world network on the right-hand side shows only one optimal node.
So, assuming that in the proposed model polled nodes are randomly selected, as well as the two sources, we implement a Monte Carlo simulation.
Notably, from a theoretical point of view, being $N^*$ the number of optimal positions in the network, the probability that a polled node occupies that position is $p(D^*) = \frac{N^*}{N}$.
Thus, increasing the amount of polled nodes, we have more chances to select one located in a convenient position.
Hence, since for each network configuration the amount of time steps required for a rapid detection has a minimum (i.e. that obtained when polled nodes are located in convenient positions), increasing the set of polled nodes the average amount of time steps gets closer to the minimum value.
\section{Results}\label{sec:results}
\begin{figure}[h!]
\centering
\includegraphics[width=0.8\textwidth]{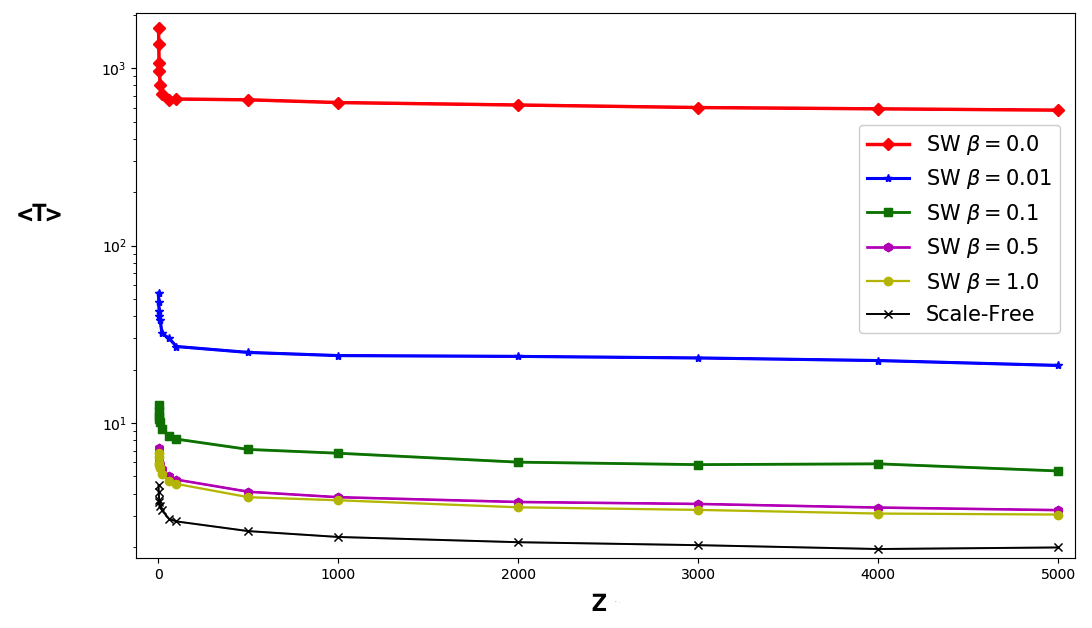}
\caption{Average number of time steps $<T>$, on varying the amount of polled nodes $Z$, for ensuring the detection of a DS attack. As reported in the legend, lines refer to the six different networks. Results have been averaged over different simulation runs. \label{fig:figure_2}}
\end{figure}

\begin{figure}[h!]
\centering
\includegraphics[width=0.8\textwidth]{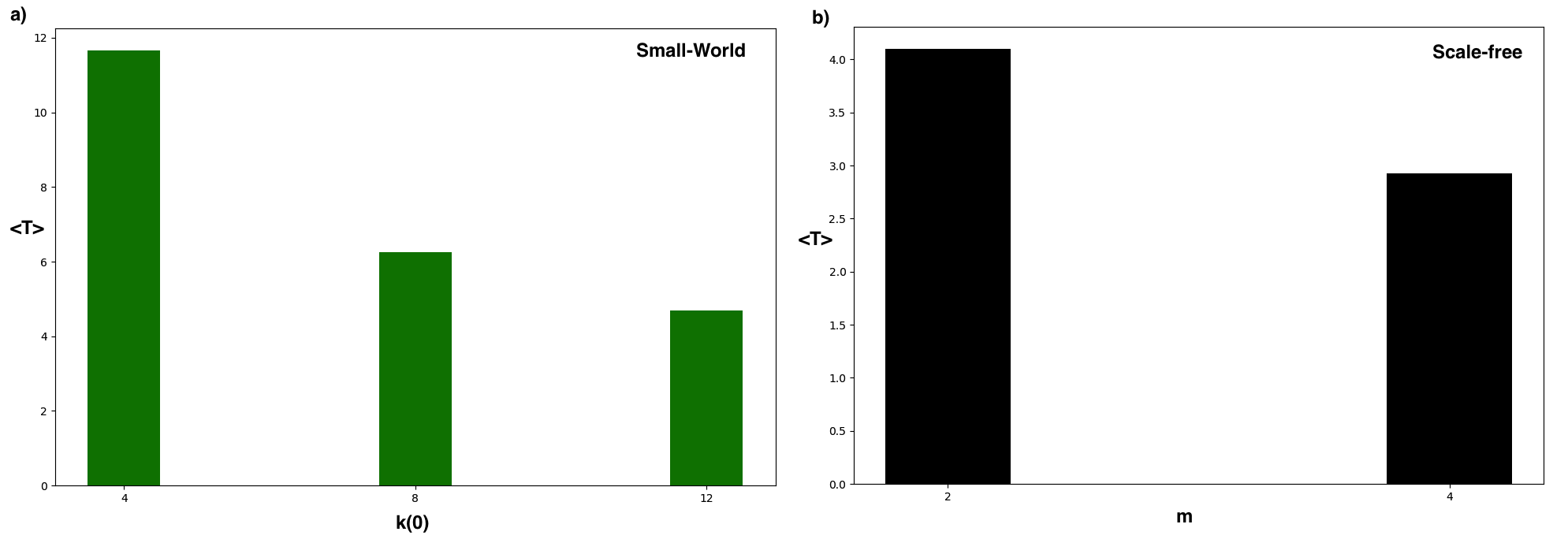}
\caption{Average number of time steps $<T>$ required for the DS detection with $Z = 2$ and $10^4$ nodes. \textbf{a)} results achieved on small-world networks obtained by $\beta = 0.1$. The label on the x-axis, i.e. $k(0)$ indicates the degree of the lattice at the beginning of its generation (i.e. before applying the re-wiring process of the Watts-Strogatz model). \textbf{b)} Results achieved on scale-free networks, with $m = 2$ and $m = 4$, i.e. on varying the minimum degree of the network used for implementing the network via the Barabasi-Albert model. Results have been averaged over different simulation runs.\label{fig:figure_4}}
\end{figure}

Numerical simulations are based on the following topologies: regular lattices, small-word networks, and scale-free networks.
In particular, the regular lattice and the small-world network have been achieved by implementing the Watts-Strogatz model~\cite{watts01}, while the scale-free configuration has been implemented via the Barabasi-Albert model~\cite{barabasi01}.
Using the Watts-Strogatz model, the connectivity pattern of the resulting network is controlled by a re-wiring parameter, usually known as $\beta$, whose range is $\beta \in [ 0,1 ]$. 
For instance, with $\beta = 0$, the network keeps its initial configuration (i.e. a regular ring lattice), whereas by increasing $\beta$ the resulting network gets provided with a small-world structure. Eventually, $\beta = 1$ leads the network to achieve a completely random structure.
Before illustrating the results, we emphasize that for each configuration we perform different simulation runs, computing both the average value (shown in following figures, and standard deviation as reported in the Table~\ref{tab:table_1}). 
In particular, for each topology, we consider $150$ different random pairs (i.e. source nodes), and for each single pair of sources we test $50$ different sets of polled nodes (randomly selected).
In addition, random topologies (i.e. small-world and scale-free) are re-generated $6$ times during each single simulation, so that the process is measured on different networks having the same pattern.
Finally, the proposed model has been studied considering networks with the following sizes: $N = [10^2, 10^3, 10^4]$. 
Figure~\ref{fig:figure_2} shows results of numerical simulations performed on networks with $N = 10^4$ nodes.
A rapid inspection of Figure~\ref{fig:figure_2} suggests that the first configuration, i.e. the regular ring lattice, requires much more time steps than all the others (in full accordance with the theoretical description before provided).
However, we decided to analyze also the worse case, i.e. setting the position of the two sources (randomly arranged in the previous simulations) in two opposite nodes of the lattice. 
For instance, labeling nodes with a number from $1$ to $10^4$, the most time consuming scenario occurs every time they have the maximum distance, e.g. one source in the node $1$ (on the top of the ring) and the other source on the node $5000$ (on the bottom of the ring). Results of this brief analysis confirm the theoretical expectation, i.e. the average amount of time steps gets closer to $1250$ as the the amount of polled nodes increases. It is worth to remind that the regular lattice here considered has an average degree equal to $4$.
Networks with a smaller amount of nodes have a behavior similar to that shown in Figure~\ref{fig:figure_2} ---see also Table~\ref{tab:table_1}.
\begin{table*}[ht]
\centering
\caption{Average amount of time steps ($<T>$), and related standard deviation (STD), for the detection of a DS attack in different networks on varying the amount of polled nodes $Z$. Results have been averaged over different simulation runs.}\label{tab:table_1}
    \begin{tabular}{ | p{2cm} | p{3cm} | p{2cm} | p{2cm} | p{2cm} |}
    \hline
    Z & Type & Nodes & $<T>$ & STD \\ \hline
    \hline
    1 & Regular & $10^3$  & 165.18 & 20.19 \\ \hline
    2 & Regular & $10^3$  & 134.05 & 22.81 \\ \hline
    10 & Regular & $10^3$  & 86.94 & 32.05 \\ \hline
    100 & Regular & $10^3$  & 69.15 & 34.17 \\ \hline
    \hline
    1 & Regular & $10^4$  & 1674.25 & 204.23 \\ \hline
    2 & Regular & $10^4$  & 1376.26 & 267.51 \\ \hline
    10 & Regular & $10^4$  & 801.98 & 251.02 \\ \hline
    100 & Regular & $10^4$  & 670.14 & 329.67 \\ \hline
    1000 & Regular & $10^4$  & 640.37 & 312.37 \\ \hline
    \hline
    \hline
    1 & SW $\beta = 0.5$ & $10^3$  & 5.17 & 0.11 \\ \hline
    2 & SW $\beta = 0.5$ & $10^3$  & 4.75 & 0.27 \\ \hline
    10 & SW $\beta = 0.5$ & $10^3$  & 3.87 & 0.43 \\ \hline
    100 & SW $\beta = 0.5$ & $10^3$  & 2.88 & 0.38 \\ \hline
    \hline
    1 & SW $\beta = 0.5$ & $10^4$  & 7.24 & 0.26 \\ \hline
    2 & SW $\beta = 0.5$ & $10^4$  & 6.79 & 0.14 \\ \hline
    10 & SW $\beta = 0.5$ & $10^4$  & 5.81 & 0.16 \\ \hline
    100 & SW $\beta = 0.5$ & $10^4$  & 4.8 & 0.22 \\ \hline
    1000 & SW $\beta = 0.5$ & $10^4$  & 3.82 & 0.41 \\ \hline
    \hline
    \hline
    1 & SF & $10^3$  & 3.52 & 0.36 \\ \hline
    2 & SF & $10^3$  & 3.14 & 0.31 \\ \hline
    10 & SF & $10^3$  & 2.51 & 0.39 \\ \hline
    100 & SF & $10^3$  & 1.89 & 0.32 \\ \hline
    \hline
    1 & SF & $10^4$  & 4.45 & 0.39 \\ \hline
    2 & SF & $10^4$  & 4.12 & 0.31 \\ \hline
    10 & SF & $10^4$  & 3.4 & 0.35 \\ \hline
    100 & SF & $10^4$  & 2.79 & 0.4 \\ \hline 
    1000 & SF & $10^4$  & 2.28 & 0.42 \\ \hline
    \end{tabular}
   
\end{table*}
It is worth to highlight that all networks show a similar behavior, i.e. the first part of each line in Figure~\ref{fig:figure_2} rapidly decreases until a kind of elbow makes it almost flat, indicating that after a given number of polled nodes (i.e. $Z$) the performance of the system becomes almost constant. So, in principle, it is possible to find an optimal $Z$ that allows to implement a reliable oracle.
Then, we study the behavior of the model, scaling the size of the networks.
\begin{figure}[h!]
\centering
\includegraphics[width=1.0\textwidth]{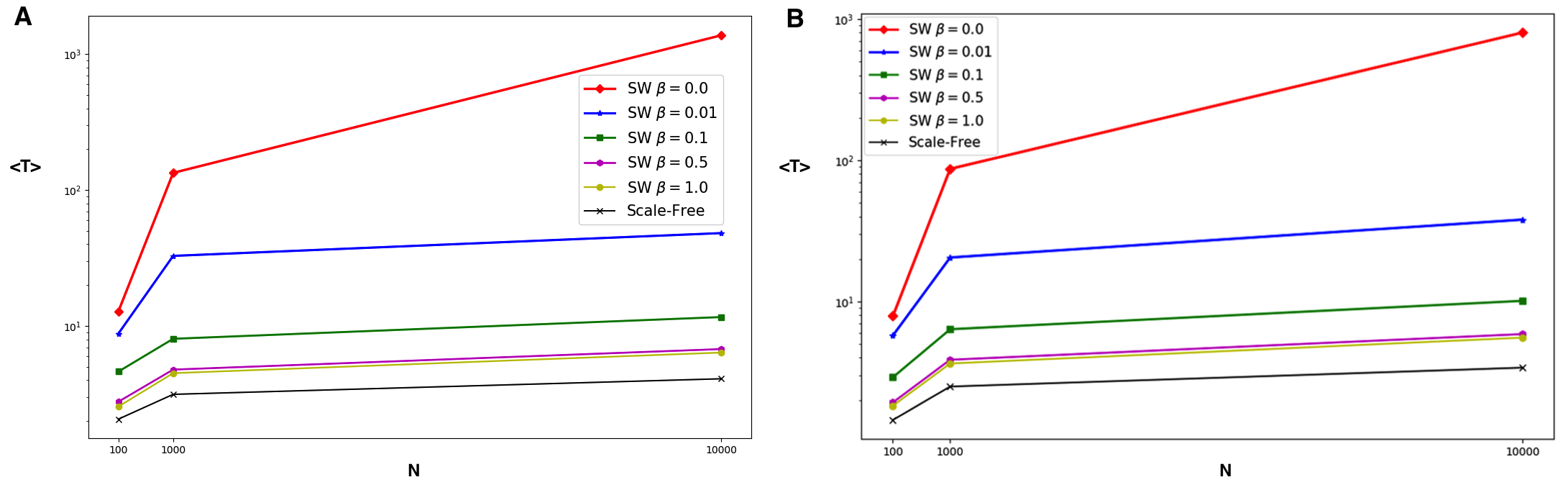}
\caption{Average amount of time steps $<T>$  for ensuring the detection of a DS attack, on varying the size $N$ of the considered networks ---see the inner legend. \textbf{A} shows results achieved by considering $Z = 2$, while \textbf{B} by considering $Z = 10$. Results have been averaged over different simulation runs.\label{fig:figure_3}}
\end{figure}
Remarkably, in random networks, few polled nodes are able to keep a good performance even if the network grows, i.e. increasing $N$.
Finally, although expected according the previous theoretical considerations, we analyze the model on networks provided with a higher density of edges. Figure~\ref{fig:figure_4} refers to small-world networks achieved by setting $\beta = 0.1$ and increasing the initial average degree (i.e. $k(0)$), and to scale-free networks implemented by setting the minimum degree $m$ to $2$ and to $4$. Both kinds of networks have $10^4$ nodes. Results confirm the theoretical hypothesis (i.e. increasing the density of edges, the value of $<T>$ decreases).
\section{Discussion and Conclusion}\label{sec:conclusion}
In this work, we investigate the dynamics of DS attacks in different complex topologies, in order to design a detection system for the Bitcoin network.
Notably, we aim to quantify the relation between the topology of a network and the amount of nodes that need to be polled by an 'oracle' devised for identifying the presence of conflicting transactions. Since it is not possible to double spend the same Bitcoin, the issue to consider is only related to possible attempts in defrauding users like merchants, willing to accept payments in Bitcoin.
Here, we assume that the Bitcoin network has a non-trivial topology (see also~\cite{javarone02}). For this reason, we deem useful to investigate this kind of fraudulent behavior by using the framework of complex networks, so that the related outcomes can constitute a reference for implementing a detection system in the real network.
In order to evaluate how random topologies affect the analyzed process, our analysis considers also regular ring graphs. Thus, the motivation of this choice goes beyond the scientific curiosity. Then, we analyze also famous topologies as small-world networks and scale-free networks. Moreover, we remind that when implementing small-world networks according to the Watts-Strogatz model (as in our case), the first step is the definition of a regular ring lattice.
Looking at results, we observe that random networks, as small-world and scale-free networks, are faster than regular graphs in the detection of DS attacks. That result, as above reported, is quite expected. Notably, the broadcasting of transactions is similar to a spreading process on networks, and the latter has been proven being faster in the above mentioned random topologies than in the those regular.
However, results show that both small-world networks and scale-free networks require to poll very few nodes for ensuring a rapid detection. In addition, it is worth to highlight that the performance of an oracle polling few nodes is poorly affected by the scaling of the network (i.e. the same amount of polled nodes can be used even when the network substantially grows).
So, we deem that the achieved outcomes can be used as reference for implementing a DS detection system on the Bitcoin network. Notably, estimating the connectivity pattern of the Bitcoin network might allow to evaluate the minimum amount of nodes when polling. In addition, the proposed model assumes that polled nodes are randomly selected so, in principle, the same approach can be applied also in the real system.
Before to conclude, we provide some further observations. First, the Bitcoin network is a direct network (i.e. the edges can be represented as arrows) instead, in the proposed model, for the sake of simplicity we considered undirected networks. 
The central core of Bitcoin miners is not represented in our model. However, we suggest that an oracle working in the Bitcoin network should perform the polling only considering miners, since they can store a very high amount of connections. 
In particular, when the network grows, their average degree increases (i.e. $\frac{d<k_m>}{dt} > 0$, with $<k_m>$ average degree of miners). It is worth to clarify that, even if we have analyzed also the case with only one polled node, in the real system at least three nodes should be selected, as only one could collude with a fraudulent user. 
Furthermore, we emphasize that when implementing a real oracle, the 'honest transaction' can be used as reference during the poll, i.e. an oracle would had only to evaluate if polled nodes received transactions that are conflicting with the honest one it knows.
Eventually, beyond the specific application of our model to the problem of conflicting transactions, we suggest that it might be exploited also for facing problems in other fields, e.g. when the topology of a network can only be inferred and a spreading process involves more than one entity.

\end{document}